\documentclass[final,1p,times,twocolumn]{elsarticle}

\usepackage{amssymb}
\usepackage{amsmath}

\usepackage{color}
\usepackage{url}
\usepackage{natbib}
\usepackage{graphicx}
\usepackage{subcaption}
\usepackage{overpic}

\usepackage{hyperref}

\newcounter{bla}

\journal{Computer Physics Communications}

\begin{document}

\begin{frontmatter}



\title{Update of PHYSBO: Improving Usability and Portability of Bayesian Optimization for Physics and Materials Research}

\author[a]{Yuichi Motoyama\corref{author}}
\author[a]{Kazuyoshi Yoshimi}
\author[a]{Tatsumi Aoyama}
\author[b]{Kei Terayama}
\author[c,d]{Koji Tsuda}
\author[c,d]{Ryo Tamura\corref{author}}

\cortext[author] {Corresponding author.\\\textit{E-mail address:} y-motoyama@issp.u-tokyo.ac.jp (Motoyama), tamura.ryo@nims.go.jp
 (Tamura)}
\address[a]{Institute for Solid State Physics, The University of Tokyo, Chiba 277-8581, Japan}
\address[b]{Graduate School of Medical Life Science, Yokohama City University, Yokohama 230-0045, Japan}
\address[c]{Graduate School of Frontier Science, The University of Tokyo, Chiba 277-8561, Japan}
\address[d]{Center for Basic Research on Materials, National Institute for Materials Science, Ibaraki 305-0044, Japan}

\begin{abstract}
Bayesian optimization (BO) is widely used to accelerate physics and materials research, where objective function evaluations are computationally or experimentally expensive. While many BO frameworks focus on algorithmic efficiency, practical usability and portability are equally critical for sustained use in real research environments. PHYSBO is a Bayesian optimization library designed to address these needs by enabling optimization over user-defined candidate pools and by supporting domain-specific problem settings.

This paper presents the major updates introduced in PHYSBO versions 2 and 3, with a focus on improvements in usability, portability, and practical deployment rather than on new optimization algorithms.
In PHYSBO version 2, the software license was changed from GPL to MPL to improve compatibility with a wider range of research and software ecosystems. Building on this revision, PHYSBO version 3 introduces a set of implementation-oriented updates aimed at improving usability and portability, without modifying the core optimization algorithms. These updates include improvements in computational performance and scalability, extended support for multi-objective optimization, the introduction of range-based policies for continuous-variable optimization, the removal of environment-dependent components such as tightly coupled Cython modules, and compatibility with NumPy 2.

These improvements reduce the technical and organizational burden on users, enabling PHYSBO to be deployed across diverse computing environments and research workflows. By emphasizing portability and ease of integration while maintaining sufficient performance, PHYSBO version 3 is positioned as a sustainable research infrastructure for Bayesian optimization in physics and materials science.
\end{abstract}


\end{frontmatter}



{\bf PROGRAM SUMMARY/NEW VERSION PROGRAM SUMMARY}

\begin{small}
\noindent
{\em Program Title:} PHYSBO \\
{\em CPC Library link to program files:} (to be added by Technical Editor) \\
{\em Developer's repository link:} \url{https://github.com/issp-center-dev/PHYSBO} \\
{\em Code Ocean capsule:} (to be added by Technical Editor) \\
{\em Licensing provisions:} MPL-2.0 \\
{\em Programming language:} Python \\
{\em Supplementary material:} None \\
{\em Journal reference of previous version:} 
Y. Motoyama et al., Computer Physics Communications 278 (2022) 108405. \\
{\em Does the new version supersede the previous version?:} Yes \\
{\em Reasons for the new version:}\\
PHYSBO version 3 was developed to improve usability, portability, and long-term sustainability in real research environments. While previous versions provided efficient Bayesian optimization for discrete candidate spaces, practical use revealed limitations related to software environment dependency, restricted support for continuous and multi-objective optimization, and licensing constraints that hindered integration into larger workflows. The new version addresses these issues to enable broader and more flexible deployment. \\
{\em Summary of revisions:}\\
Major updates in PHYSBO version 3 include improvements in computational performance and scalability, enhanced support for multi-objective optimization, and the introduction of range-based policies for continuous-variable optimization. Environment-dependent components such as tightly coupled Cython modules have been removed, and compatibility with NumPy 2 has been added. In addition, the software license has been changed from GPL to MPL to facilitate easier integration into diverse research workflows. \\
{\em Nature of problem (approx. 50--250 words):}\\
Optimization problems in physics and materials science often involve expensive objective function evaluations, such as first-principles calculations, large-scale simulations, or experimental measurements. Bayesian optimization is a powerful approach for reducing the number of evaluations required to identify optimal parameters. However, practical application of Bayesian optimization in research settings faces several challenges, including scalability to large search spaces, handling of multiple competing objectives, representation of continuous design variables, and software environment compatibility. In addition, researchers often need to incorporate domain-specific constraints and prior knowledge into the optimization process, which is not always supported by generic optimization frameworks. \\
{\em Solution method (approx. 50--250 words):}\\
PHYSBO provides a Bayesian optimization framework tailored to physics and materials science applications. It supports optimization over user-defined candidate pools as well as range-based policies that enable direct optimization over continuous variables. PHYSBO version 3 extends support for multi-objective optimization using established approaches such as scalarization and non-dominated sorting. The software is designed to balance computational efficiency with practical usability by minimizing environment-dependent components and ensuring compatibility with modern numerical libraries. These design choices allow PHYSBO to be easily integrated into simulation and experimental workflows while maintaining sufficient performance for real-world optimization tasks. \\
{\em Additional comments including restrictions and unusual features (approx. 50--250 words):}\\
PHYSBO is intended for use as a research library and can be imported into larger software frameworks and workflows. The adoption of the MPL license in version 3 reduces potential legal barriers associated with library usage and facilitates industry--academia collaboration. Performance-critical components are implemented to ensure practical efficiency, while prioritizing portability and ease of installation. PHYSBO does not impose restrictions on the form of objective functions, allowing users to integrate arbitrary external simulation or experimental codes.
\end{small}
   \\

\section{Introduction}
Bayesian optimization (BO) has become an essential tool for accelerating materials and physics research, where objective function evaluations are often expensive and time-consuming[cite]. Typical applications include parameter optimization for first-principles calculations and large-scale simulations, as well as experimental design, where each evaluation entails substantial computational or experimental costs. In such settings, the practical usability of BO libraries is as important as their algorithmic performance.

PHYSBO is a Bayesian optimization (BO) library engineered for applications in physics and materials science\cite{Motoyama2022}.
A defining feature of PHYSBO is its ability to perform optimization over a predefined, discrete candidate pool, allowing researchers to seamlessly incorporate domain-specific knowledge, such as chemical intuition, physical constraints, and experimental limitations, into the exploration process.
While this design is particularly advantageous for physics and materials science, the underlying framework is not restricted to these domains and can be applied more broadly to discrete and constrained optimization problems in various scientific and engineering fields, where evaluations are costly and prior knowledge is essential.
This flexibility distinguishes it from generic BO frameworks that often struggle to integrate such nuanced priors.
Furthermore, PHYSBO is designed to be virtually parameter-free, minimizing the need for manual hyperparameter tuning to ensure better performance across diverse, real-world problems.

The library’s utility is evidenced by its successful application in various areas, ranging from simulation-based atomic configuration exploration\cite{2511.17972} and the discovery of next-generation magnetic materials\cite{Nakamura2025} to the design of thermal metamaterials\cite{Xi2023}.
redIts efficacy extends to a broad range of experimental domains as well, notably in the development of electrocatalysts\cite{Sakaushi2023}, superconducting materials\cite{Yamazaki2025}, polymer composites\cite{Xu2025,amamoto2025machine}, drug discovery\cite{yoshizawa2025data}, antimicrobial peptides\cite{murakami2023design}, quantum chemistry calculations\cite{terayama2023koopmans}, estimation of exciton Hamiltonian\cite{Shimooka2025}, and fish feed\cite{terayama2025data}.
Beyond these specific applications, PHYSBO has also been integrated into general-purpose data-driven frameworks such as 2DMAT\cite{MOTOYAMA2022108465}, where it is employed for efficient parameter optimization in experimental and data-analysis workflows.
More recently, as the paradigm of self-driving laboratories\cite{doi:10.1126/sciadv.aaz8867,Tom2024,Yoshikawa2025} gains momentum, PHYSBO has been adopted as the core optimization engine within the NIMO library\cite{Tamura2023}, serving as the ``algorithmic brain'' for various autonomous experimental systems\cite{Matsuda2019,Akiyama2024,Tamura2025,Toyama2025}.
Furthermore, PHYSBO has been explored in community-driven events such as the AIMHack2024 hackathon, where researchers from materials science, information science, and related fields engaged in collaborative projects involving data-driven optimization workflows, highlighting its accessibility and practical utility beyond conventional research settings\cite{AIMHack2025}.

However, as PHYSBO has been increasingly used in real research environments, several limitations have become apparent. 
The current framework faces significant technical and strategic challenges, including a heavy reliance on specific environments like Cython and scalability bottlenecks during large-scale multi-objective optimization. While PHYSBO was originally centered on the paradigm of candidate pool-based optimization, there is an increasing demand for a unified tool capable of handling continuous design variables, thereby eliminating the operational overhead of switching to other software such as GPyOpt or Optuna. Furthermore, long-term sustainability and broader adoption have been constrained by compatibility issues with evolving numerical libraries and restrictive licensing conditions that limit its utility in industrial and collaborative environments.
PHYSBO version 3 was developed to address these challenges, with a primary focus on improving usability, portability, and practical deployability, rather than introducing fundamentally new optimization algorithms. The updates include improvements in computational performance and scalability, enhanced support for multi-objective optimization, the introduction of range-based policies for continuous-variable optimization, the removal of environment-dependent components such as tightly coupled Cython modules, compatibility with NumPy 2, and a change of software license from GPL to MPL.

In this paper, we present an overview of the major updates introduced in PHYSBO version 3 (3.2.0 released on 2026-02-06) and discuss their impact on practical research workflows. Rather than focusing solely on algorithmic details, we emphasize how these changes reduce the burden on users and enable more flexible and sustainable use of BO in real-world physics and materials research.

\section{License Change: From GPL to MPL}
Software licensing plays an important role in determining how research software can be used, integrated, and redistributed in practical research environments. Earlier versions of PHYSBO version 1 were distributed under the GNU General Public License (GPL), which ensures strong copyleft protection but can introduce limitations when the software is combined with other codes or libraries.
In particular, the use of GPL-licensed software may cause concerns when PHYSBO is imported as a library into larger software frameworks or workflows. In some cases, simply importing a GPL-licensed package can raise legal or organizational issues, especially in collaborative projects involving industrial partners or mixed-license codebases. These concerns can hinder the adoption of otherwise useful research software, even when it is used purely as a dependency rather than as a standalone application.

To address these issues, PHYSBO version 2 adopted the Mozilla Public License (MPL). MPL provides a file-level copyleft mechanism that preserves openness of modifications to PHYSBO itself, while allowing the library to be imported and used within broader software ecosystems under different licenses. This change significantly improves compatibility with other scientific software and reduces potential legal barriers associated with library usage. The license change reflects a broader shift in the design philosophy of PHYSBO toward practical and sustainable deployment as a research infrastructure.
\section{Software Portability: Removing Environment Dependency}

In practical research environments, software portability and ease of installation are critical factors that strongly affect the usability of numerical optimization libraries.
In earlier versions of PHYSBO, performance-critical components were implemented using Cython to accelerate Gaussian process--based computations.
While this approach reduced computational overhead, it also introduced significant environment dependencies.
In particular, building Cython extensions required specific compiler settings and library versions, which often caused installation difficulties on certain platforms, including Windows systems and heterogeneous high-performance computing environments.

In PHYSBO version~3, this tight coupling between the core library and Cython-based implementations has been removed.
The core functionality of PHYSBO is now provided as a pure Python package by default, eliminating the need for platform-specific compilation steps.
As a result, installation is significantly simplified, and the library can be used more reliably across diverse research environments.
The performance-critical parts previously implemented with Cython are further revised using the NunPy arithmetic operations to maintain comparable execution times.
Therefore, this design change does not lead to a practical degradation in performance for typical Bayesian optimization workflows.
In many realistic use cases, the overall optimization cost is dominated by
objective function evaluations rather than surrogate model updates, making the
pure Python implementation sufficient for practical applications.

Another important update in PHYSBO version~3 is compatibility with NumPy~2.0 and the later versions.
As the Python scientific ecosystem evolves, maintaining compatibility with
modern numerical libraries is essential for long-term sustainability.
By supporting NumPy~2.0, PHYSBO can be seamlessly integrated into contemporary
Python environments without relying on legacy configurations.

\section{New Multi-objective Optimization methods}

\subsection{Multi-objective Optimization Problem and Scalarization of them}

When multiple objective functions are to be optimized simultaneously, it is generally impossible to improve all objectives at once, and trade-offs among them inevitably arise.
Such trade-offs are characterized by the concept of the Pareto front.

Let $\boldsymbol{y}_i = (y_{i,1}, \dots, y_{i,p})$ and $\boldsymbol{y}_j = (y_{j,1}, \dots, y_{j,p})$ denote objective vectors for two candidate points $\boldsymbol{x}_i$ and $\boldsymbol{x}_j$, where all objectives are assumed to be maximized.
We say that $\boldsymbol{y}_i$ dominates $\boldsymbol{y}_j$ if
\begin{equation}
\forall a \,\,  y_{i,a} \ge y_{j,a} \quad \text{and} \quad \exists a \,\, y_{i,a} > y_{j,a}.
\end{equation}

A candidate solution is said to be \emph{Pareto-optimal} if it is not dominated by any other solution in the dataset. The set of all Pareto-optimal solutions forms the \emph{Pareto front}, which represents the optimal trade-offs among competing objectives.
Given a Pareto front $\{\boldsymbol{y}_i^{\mathrm{P}}\}$ and a reference point $\boldsymbol{y}^{\mathrm{ref}}$, the dominated region is defined as the union of hyperrectangles
\begin{equation}
\Omega_i = \prod_{a=1}^p \left[ y_a^{\mathrm{ref}}, \, y_{i,a}^{\mathrm{P}} \right].
\end{equation}
The hypervolume of the dominated region is commonly used as a performance metric to quantitatively assess the quality of a Pareto front.

In previous versions of PHYSBO, a separate surrogate model is constructed for each objective, and candidate points are selected by maximizing acquisition functions based on the hypervolume, such as the hypervolume probability of improvement (HVPI) and the expected hypervolume improvement (EHVI).
However, the computational cost of hypervolume calculation grows exponentially with the number of objectives, which can become a significant bottleneck in practical applications. To address this issue, an alternative approach is to scalarize multiple objectives into a single objective function $z_i$ and perform Bayesian optimization on this transformed objective.
PHYSBO version 3 implements another \texttt{Policy} class based on this approach, \texttt{physbo.search.discrete\_unified.Policy}.
This class has a similar interface with the existing multi-objective policy, \texttt{physbo.search.discrete\_multi.Policy}.
The main difference is that the \texttt{discrete\_unified.Policy.bayes\_search} method takes an extra argument, \texttt{unify\_method}, to specify the scalarization strategies for calculating $z_i$ from $y_{i,a}$.
In the present version of PHYSBO, we implement two such scalarization strategies: ParEGO and non-dominated sorting (NDS).

\subsection{ParEGO}
ParEGO (Pareto Efficient Global Optimization) is a scalarization-based approach for multi-objective Bayesian optimization, originally proposed by Knowles~\cite{Knowles2006}. The key idea is to transform a $p$-objective optimization problem into a sequence of single-objective problems by means of randomized scalarization.

Let $\boldsymbol{y}_i = (y_{i,1}, y_{i,2}, \dots, y_{i,p})$ denote the vector of objective values evaluated at design point $\boldsymbol{x}_i$.
ParEGO constructs a scalar objective function $z_i$ from $\boldsymbol{y}_i$ as
\begin{equation}
z_i
= \rho_{\mathrm{sum}} \sum_{a=1}^{p} w_a y_{i,a}
+ \rho_{\mathrm{max}} \max_{a} \left( w_a y_{i,a} \right),
\label{eq:parego_scalarization}
\end{equation}
where $\boldsymbol{w} = (w_1, \dots, w_p)$ is a weight vector satisfying
$\sum_a w_a = 1$, and $\rho_{\mathrm{sum}}$ and $\rho_{\mathrm{max}}$ are hyperparameters that control the relative contributions of the weighted sum term and the weighted Chebyshev term, respectively.

In PHYSBO, the weight vector $\boldsymbol{w}$ is either randomly sampled at each Bayesian optimization step or generated from a discretized simplex.
Each objective $y_{i,a}$ is normalized using the minimum and maximum values observed in the current training data, so that all objectives contribute on a comparable scale.
The resulting scalar objective $z_i$ is then modeled using a Gaussian process regression, and standard acquisition functions such as expected improvement (EI)
are applied.
By repeatedly changing the weight vector $\boldsymbol{w}$ across optimization steps, ParEGO implicitly explores different trade-offs among objectives and enables efficient approximation of the Pareto front, while avoiding the high computational cost associated with directly optimizing multi-objective acquisition functions such as hypervolume improvement.

In PHYSBO version~3, ParEGO is implemented as \texttt{physbo.search.unify.ParEGO}, and can be used through the unified multi-objective policy without modifying the underlying Bayesian optimization workflow.

\subsection{NDS}

Non-dominated sorting (NDS) is a fundamental technique in multi-objective optimization for classifying candidate solutions according to Pareto dominance relations.
Unlike ParEGO, which relies on randomized scalarization of objective values, NDS does not introduce additional scalarization weights.
Instead, it directly exploits dominance relations among objective vectors, making it particularly suitable for post-analysis and interpretation of multi-objective optimization results.
NDS assigns a rank $r_i$ to each solution $i$ by recursively identifying Pareto fronts.
All non-dominated points in the current dataset are assigned rank $r_i = 1$, corresponding to Pareto-optimal solutions.
After removing these points, the non-dominated points of the remaining set are assigned rank $r_i = 2$. This procedure is iterated with increasing rank values $r_i = 3, 4, \dots$ until the entire set of candidate solutions is exhaustively partitioned. In this manner, NDS decomposes the solution set into a sequence of Pareto fronts and provides an ordinal measure of the relative quality of candidate solutions.
The rank obtained by NDS is converted into a scalar objective value as $z_i = 1/r_i$,
which is maximized in the subsequent Bayesian optimization procedure.
In this formulation, solutions on better Pareto fronts are assigned larger objective values, allowing multi-objective optimization to be reduced to a single-objective problem while preserving Pareto dominance information.

In PHYSBO version~3, NDS is implemented as \texttt{physbo.search.unify.NDS}, where the maximum rank is used for scalarization, can be controlled by the user. For points $i$ with larger rank, $z_i$ will be $0$.

\subsection{Numerical example}

In the following, we present benchmark results comparing ParEGO and NDS with an existing method, HVPI, using three test functions: VLMOP2, KYMN, and Gaussians.
\footnote{
All the scripts used in this paper are available from ISSP data repository~\cite{PHYSBO_gallery}.
}
\footnote{
These benchmarks were performed on Supercomputer System C (kugui) of the Institute for Solid State Physics, the University of Tokyo (AMD EPYC 7763, 2.45GHz, $64 \times 2$cores).
}

\subsubsection{VLMOP2 benchmark}

\begin{figure}[t]
  \centering
  \includegraphics[width=0.9\linewidth]{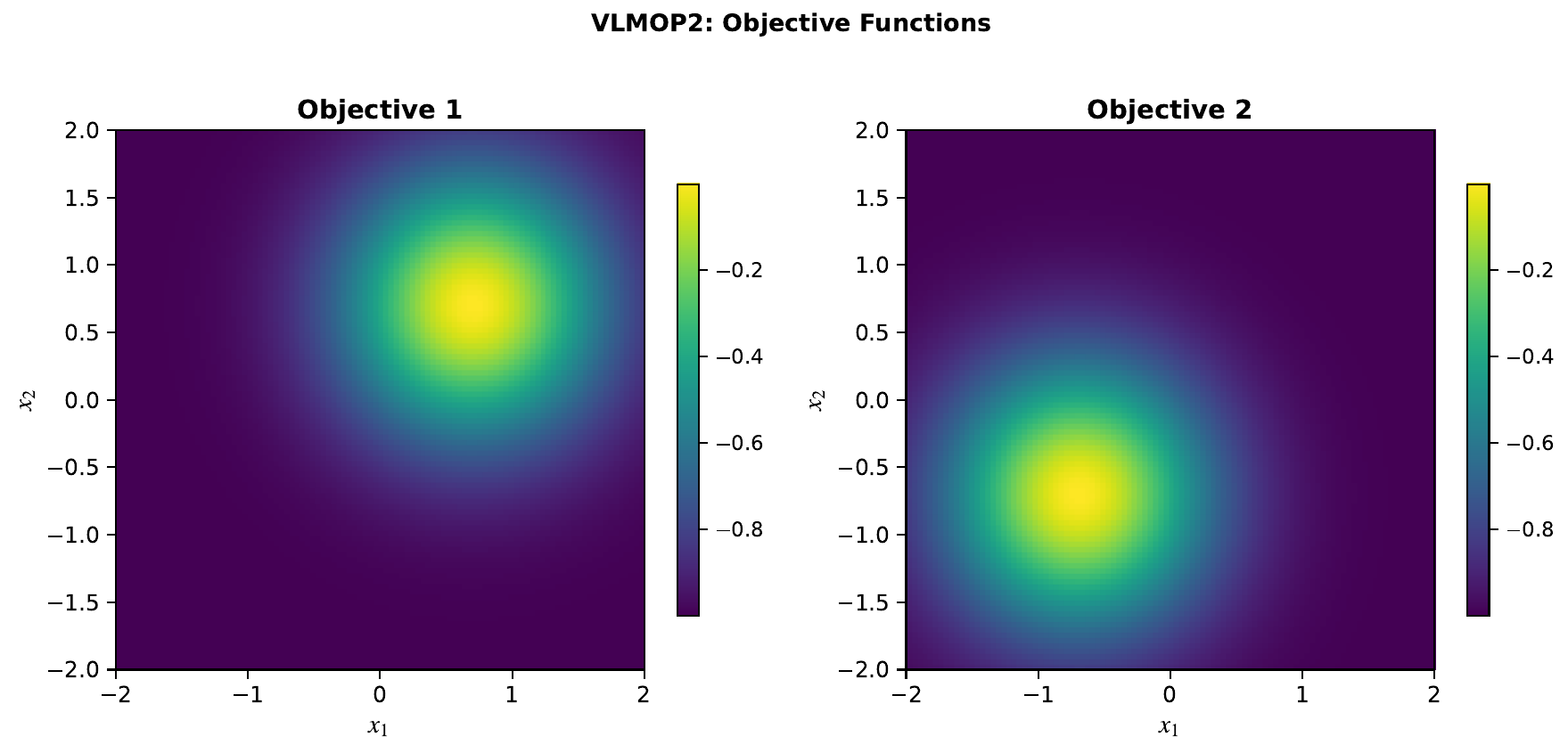}
  \caption{Objective function landscapes of the VLMOP2 benchmark.
  Left: $-y_1(\boldsymbol{x})$.
  Right: $-y_2(\boldsymbol{x})$.
  The two objectives attain their maxima at different locations in the design space, inducing a trade-off between them.}
  \label{fig:vlmop2_objectives}
\end{figure}

We first consider the VLMOP2 benchmark problem\cite{Fonseca1995, vanVeldhuizen1999}, which is widely used to evaluate multi-objective optimization algorithms.
In this problem, a solution $\boldsymbol{x} = (x_1, x_2) \in \mathbb{R}^2$ is evaluated by two objective functions
\begin{align}
y_1(\boldsymbol{x})
&= 1 - \exp\left[-\sum_{i=1}^2 \left(x_i - \frac{1}{\sqrt{2}}\right)^2\right], \\
y_2(\boldsymbol{x})
&= 1 - \exp\left[-\sum_{i=1}^2 \left(x_i + \frac{1}{\sqrt{2}}\right)^2\right].
\end{align}
These objectives attain their minimum values of $0$ at
$\boldsymbol{x} = (\pm 1/\sqrt{2}, \pm 1/\sqrt{2})$, respectively, and are bounded above by $1$.
Since PHYSBO assumes a maximization setting, the optimization is performed on the negated objectives,
\[
\tilde{\boldsymbol{y}}(\boldsymbol{x})
= \bigl(-y_1(\boldsymbol{x}), -y_2(\boldsymbol{x})\bigr),
\]
so that larger values correspond to Pareto-improving solutions.
Figure~\ref{fig:vlmop2_objectives} visualizes the two objective functions over the two-dimensional design space $[-2,2]\times[-2,2]$.
Each objective exhibits a single smooth basin with its maximum located near $(\pm 1/\sqrt{2}, \pm 1/\sqrt{2})$, and the two optima are spatially separated.
As a result, improving one objective necessarily degrades the other, giving rise to a smooth and continuous Pareto front.

\begin{figure}[t]
  \centering
  \includegraphics[width=0.95\linewidth]{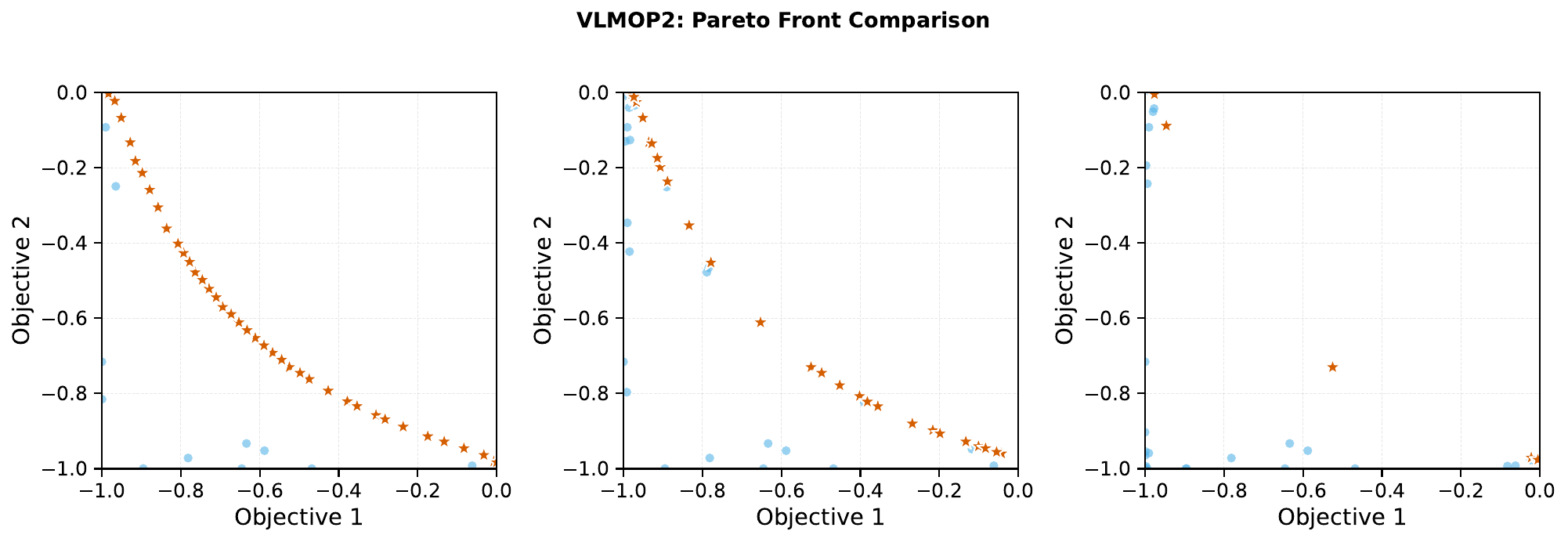}
  \caption{Pareto front comparison for the VLMOP2 benchmark.
  Blue points denote evaluated solutions, while orange markers indicate the
  non-dominated solutions.
  HVPI yields the most complete coverage of the Pareto front, NDS provides a
  competitive approximation at a significantly lower computational cost, and
  ParEGO produces a sparse set of trade-off solutions.}
  \label{fig:vlmop2_pareto}
\end{figure}

The input space is discretized on a two-dimensional domain $[-2,2]\times[-2,2]$, and a Cartesian grid with $101$ points per dimension is used, yielding $10{,}201$ candidate solutions.
Bayesian optimization is carried out with an initial $10$ random sampling steps\footnote{It should be noted that common random samples are used for all three methods.} followed by $40$ optimization steps using either HVPI, NDS-based unification with $r_\text{max} = 10$, or ParEGO-based scalarization with $\rho_\text{max}=1$ and $\rho_\text{sum} = 0.05$.
Figure~\ref{fig:vlmop2_pareto} compares the Pareto fronts obtained by the three methods.
HVPI achieves the largest dominated hypervolume\footnote{The reference point of the dominated region is $\boldsymbol{y}^\text{ref} = (-1, -1)$.} (HV $=0.3285$), indicating the most extensive coverage of the smooth and continuous Pareto front.
However, this performance comes at the cost of the longest computation time. In contrast, the NDS-based approach attains a comparable hypervolume
(HV $=0.3092$) while reducing the computational time by more than a factor of two. ParEGO exhibits the shortest computation time, but its Pareto front is sparse and fails to capture the continuous trade-off curve, resulting in a substantially smaller hypervolume (HV $=0.1800$).
These results indicate that, for problems with smooth and continuous Pareto fronts, NDS provides an efficient alternative to HVPI by balancing solution quality and computational cost, whereas ParEGO may suffer from limited coverage due to its scalarization-based search strategy.

\subsubsection{Kita--Yabumoto--Mori--Nishikawa benchmark}

\begin{figure}[t]
  \centering
  \includegraphics[width=0.9\linewidth]{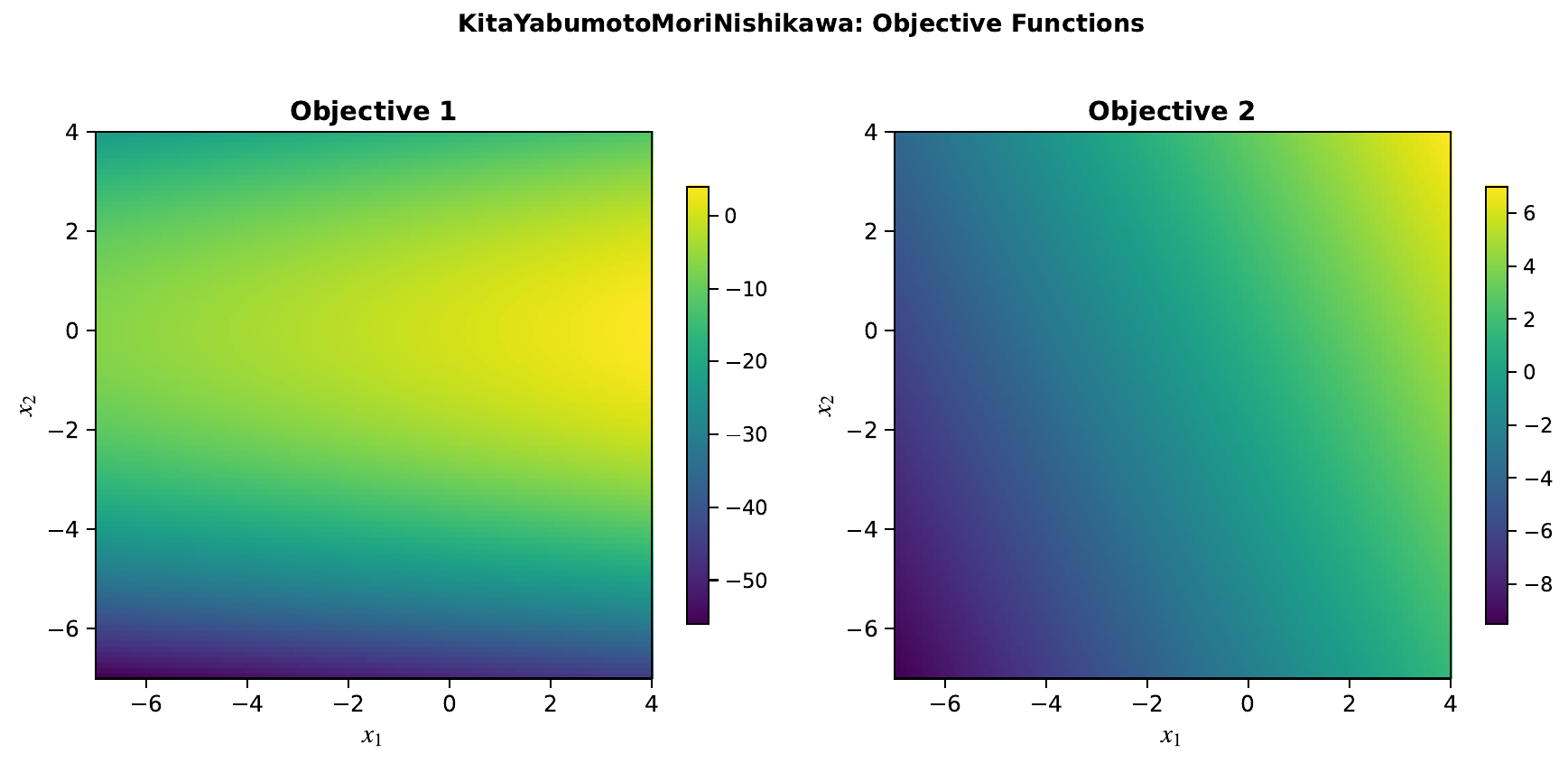}
  \caption{Objective function landscapes of the Kita--Yabumoto--Mori--Nishikawa (KYMN) benchmark.
  Left: $y_1(x_1, x_2)$.
  Right: $y_2(x_1, x_2)$.
  The feasible region is implicitly restricted by nonlinear inequality constraints, leading to a highly constrained optimization landscape.}
  \label{fig:kymn_objectives}
\end{figure}

We next consider the Kita--Yabumoto--Mori--Nishikawa (KYMN) benchmark problem\cite{Kita1996}, which is a constrained bi-objective optimization problem originally proposed to test multi-objective algorithms under nonlinear feasibility conditions.
In this problem, a solution $\boldsymbol{x} = (x_1, x_2) \in \mathbb{R}^2$ is evaluated by two objective functions
\begin{align}
y_1(x_1, x_2) &= -x_1^2 + x_2^2, \\
y_2(x_1, x_2) &= 0.5 x_1 + x_2 + 1,
\end{align}
subject to the inequality constraints
\begin{align}
g_1(x_1, x_2) &= 6.5 - \frac{x_1}{6} - x_2 \ge 0, \\
g_2(x_1, x_2) &= 7.5 - \frac{x_1}{2} - x_2 \ge 0, \\
g_3(x_1, x_2) &= 30 - 5 x_1 - x_2 \ge 0.
\end{align}
All objectives are maximized within the feasible region defined by these constraints. Figure~\ref{fig:kymn_objectives} visualizes the two objective functions over the design space.
Unlike the VLMOP2 benchmark, both objective functions vary almost monotonically along different directions in the design space, and the feasible region is strongly restricted by the constraints.
As a consequence, the set of feasible Pareto-optimal solutions occupies a narrow and irregular region in the objective space.

\begin{figure}[t]
  \centering
  \includegraphics[width=0.95\linewidth]{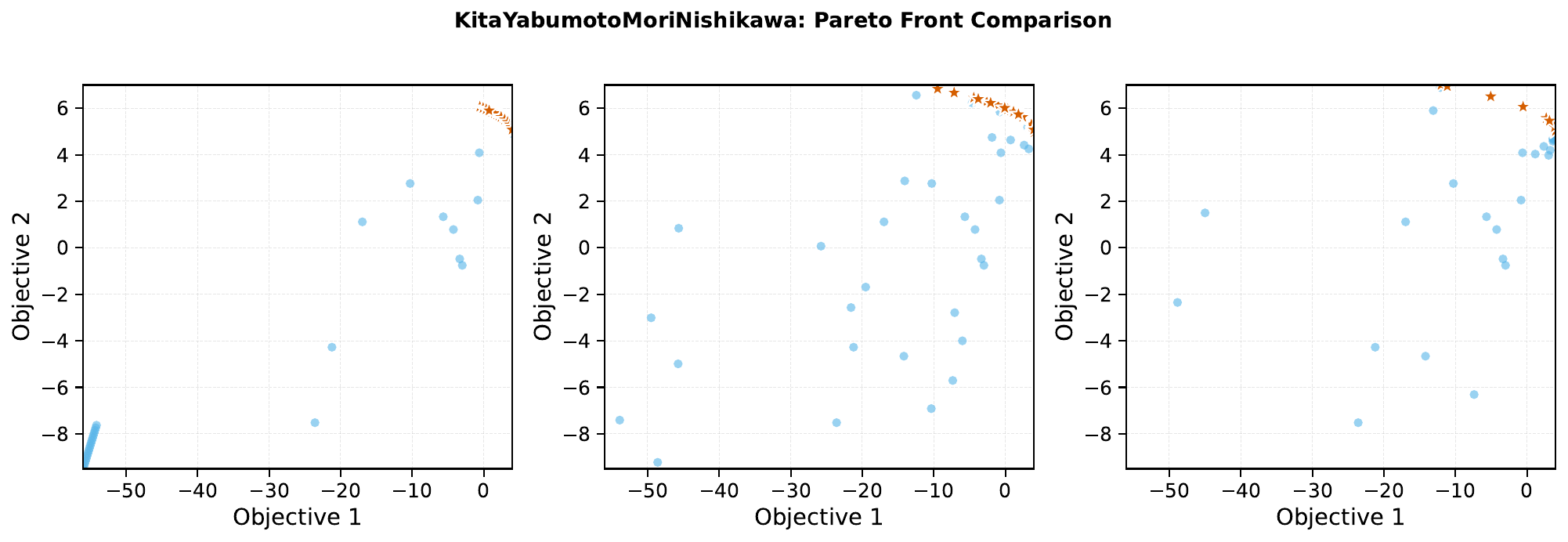}
  \caption{Pareto front comparison for the Kita--Yabumoto--Mori--Nishikawa benchmark. Blue points denote evaluated solutions, and orange markers indicate the non-dominated solutions. ParEGO concentrates sampling on a narrow feasible Pareto-optimal region, whereas NDS yields a broader distribution of near-Pareto solutions. HVPI exhibits higher computational cost and reduced coverage in this constrained scenario.}
  \label{fig:kymn_pareto}
\end{figure}

The candidate set is generated by discretizing the design space $[-7, 4] \times [-7,4]$ on a two-dimensional grid, while excluding points that violate the constraints.
Bayesian optimization is performed using the same protocol as in the VLMOP2 experiments, namely an initial $10$ random sampling steps followed by $40$ optimization steps based on either HVPI, NDS-based unification, or ParEGO-based scalarization.
Figure~\ref{fig:kymn_pareto} compares the Pareto fronts obtained by the three methods.
In contrast to the smooth Pareto front of VLMOP2, the KYMN benchmark yields a fragmented and highly nonconvex Pareto front.
In this constrained setting, both NDS and ParEGO outperform HVPI in terms of the dominated hypervolume\footnote{The reference point of the dominated region is $\boldsymbol{y}^\text{ref} = (-56.0, -9.5)$.}.
ParEGO achieves the largest hypervolume (HV $=975.9$), closely followed by NDS (HV $=970.8$), while HVPI yields a smaller value (HV $=932.1$).
From a computational perspective, both NDS and ParEGO require less than half the runtime of HVPI.
These results demonstrate that, for constrained problems with irregular and nonconvex Pareto fronts, unification-based approaches such as NDS and ParEGO can provide more efficient and effective exploration than direct hypervolume-based optimization.

Together, the VLMOP2 and KYMN benchmarks highlight how the structure of the objective landscape and the presence of constraints critically affect the relative performance of multi-objective optimization strategies.

\subsubsection{Computational time}

Finally, we consider a set of Gaussian objective functions to examine how the computational cost scales with the number of objectives.
The optimization problem is defined as
\begin{equation}
    y_a(\boldsymbol{x}) = \exp\left[-\frac{\|\boldsymbol{x} - \boldsymbol{\mu}_a\|_2^2}{2\sigma_a^2}\right],
\end{equation}
where $\boldsymbol{\mu}_a = \left[\cos\theta_a, \sin\theta_a\right]^{\mathrm{T}}$ denotes the center and $\sigma_a$ the width of the $a$-th objective function.
In this benchmark, the centers and widths are randomly generated as
\begin{align}
    \theta_a &\sim U(0, 2\pi), \\
    \sigma_a &\sim U(0.5, 1.0),
\end{align}
where $U(l,h)$ denotes the uniform distribution over the semi-closed interval $[l,h)$.
The input space $[-2,2] \times [-2,2]$ is discretized into $101 \times 101$ grid points.

As in the previous benchmarks, we first perform 10 random sampling steps for initialization and then measure the computational time over 40 Bayesian optimization steps.
To reduce the effect of the computational environment and random seeds, we conduct 10 independent trials, in which the centers and widths are regenerated.
For a fair comparison among the three methods, the centers $\{\boldsymbol{\mu}_a\}$, widths $\{\sigma_a\}$, and the points $\{\boldsymbol{x}_i\}$ selected in the initial random sampling are identical for all methods within each trial.

\begin{figure}[t]
  \centering
  \includegraphics[width=0.45\linewidth]{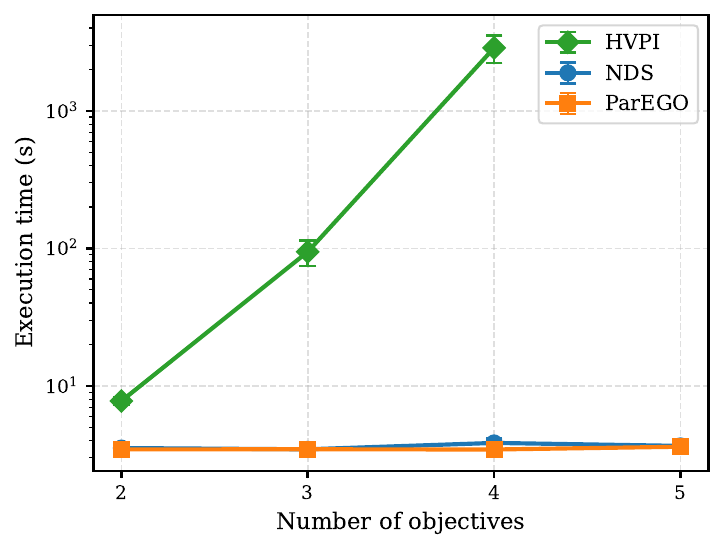}
  \includegraphics[width=0.45\linewidth]{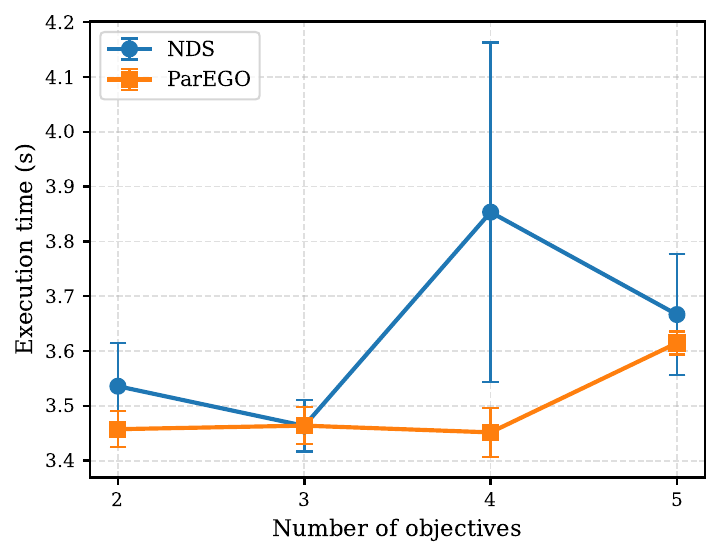}
  \caption{
  Dependence of the computational time for 40 Bayesian optimization steps on the number of objective functions $p$ for each method.
  The mean and standard deviation over 10 independent runs are shown.
  }
  \label{fig:gaussian_time}
\end{figure}

Figure~\ref{fig:gaussian_time} shows the mean and standard deviation of the computational time over 10 independent trials.
As the number of objective functions increases, the computational costs of NDS and ParEGO remain almost constant, whereas that of HVPI grows exponentially.
This exponential growth mainly originates from the computation of the hypervolume of the dominated region, which is required in the HVPI and EHVI methods.

\section{Range Policy}
\label{sec:range_policy}

\subsection{Motivation and Design Concept}
In many optimization problems encountered in physics and materials science, decision variables are naturally expressed as continuous parameters rather than elements of a finite, predefined candidate pool.
While earlier versions of PHYSBO focused on optimization over discrete candidate sets, such an approach can impose a significant burden on users when the search space is high-dimensional or inherently continuous.

To address this limitation, PHYSBO introduces the \emph{range policy}, which provides a unified framework for exploring continuous parameter spaces defined by lower and upper bounds. 
The range policy is initialized by explicitly providing these bounds, for example, as
\begin{verbatim}
policy = physbo.search.range.Policy(min_X=min_X, max_X=max_X)
\end{verbatim}
where \texttt{min\_X} and \texttt{max\_X} have the same dimensionality and collectively define a hyper-rectangular domain in the continuous parameter space.
Instead of requiring users to discretize the search space in advance, the range policy allows optimization problems to be formulated directly in terms of parameter ranges.
This design significantly reduces problem setup complexity and aligns more naturally with typical formulations of simulation and experimental parameters.

In Bayesian optimization, the next candidate point is proposed by maximizing an acquisition function, which itself constitutes an auxiliary maximization problem that must be solved at every iteration.
In the case of a discrete search space, this problem can be solved straightforwardly by evaluating the acquisition function over all remaining points in the candidate pool and selecting the maximizer. 
In contrast, for a continuous search space, an additional numerical optimization solver is required to maximize the acquisition function.

The \texttt{range.Policy.bayes\_search} function in PHYSBO is designed to accept a user-specified optimization routine for acquisition function maximization, allowing the optimization strategy to be flexibly chosen.
The current version of PHYSBO provides some optimizers as described in the following subsections.
For example, the optimizer by random sampling is initialized as
\begin{verbatim}
optimizer = physbo.search.optimize.random.Optimizer(min_X, max_X, nsamples)
\end{verbatim}
and passed to \texttt{bayes\_search} as
\begin{verbatim}
res = policy.bayes_search(optimizer=optimizer)
\end{verbatim}

As well as the single-objective problem, the multi-objective problem version of the range solver are implemented as \texttt{physbo.search.range\_multi.Policy} and \texttt{physbo.search.range\_unified.Policy}.

The range policy enhances the usability of PHYSBO by decoupling the definition of the search space from the choice of optimization algorithm. Users can define parameter ranges once and flexibly apply different search strategies without restructuring their workflows.

\subsection{Random Sampling}

In PHYSBO, random search is employed as the default strategy for acquisition function optimization: a fixed number of candidate points is sampled uniformly at random within the predefined parameter ranges, and the point that maximizes the acquisition function value is selected.
By default, the number of random samples used for acquisition optimization is set to \texttt{nsamples = 1000}, which provides a balance between robustness and computational cost. The value of \texttt{nsamples} can be adjusted by explicitly specifying a random optimizer via \texttt{physbo.search.optimizer.random.Optimizer} and passing it to the \texttt{optimizer} keyword of the \texttt{bayes\_search} function. This configurability allows users to tune the trade-off between exploration accuracy and computational overhead according to the problem size and available computational resources.

As an initialization step for optimization under the range policy, the default exploration strategy is random sampling within the specified parameter ranges.
Users define the continuous search space by specifying the lower and upper bounds for each parameter through the vectors \texttt{min\_X} and \texttt{max\_X}.
Each element of these vectors corresponds to one dimension of the search space and represents the minimum and maximum allowable values of the associated parameter.

Random sampling provides an efficient and unbiased initial exploration when little prior information about the objective function is available, while ensuring broad coverage of the search space with minimal user intervention.
This choice reflects common practices in the design of experiments, where an initial random or space-filling sampling phase is used to gather information before applying more sophisticated optimization methods.

\subsection{A wrapper of another optimization solver library, ODAT-SE}

ODAT-SE (Open Data Analysis Tool for Science and Engineering)~\cite{ODATSE_github, MOTOYAMA2022108465} is a python framework for solving optimization problems with several algorithms:
\begin{itemize}
  \item \texttt{exchange}: replica exchange Monte Carlo, effective for rugged or multimodal landscapes~\cite{HukushimaNemoto1996}.
  \item \texttt{pamc}: population annealing Monte Carlo, enabling efficient exploration of complex parameter spaces~\cite{Hukushima2003}.
  \item \texttt{minsearch}: the Nelder--Mead simplex method, suitable for smooth objective functions~\cite{NelderMead1965, Wright1996}.
  \item \texttt{mapper}: grid-based search providing systematic coverage and high reproducibility.
  \item \texttt{bayes}: Bayesian optimization based on surrogate modeling and acquisition functions.
\end{itemize}

PHYSBO provides an optimizer class wrapping ODAT-SE as \verb|physbo.search.optimize.odatse.Optimizer|.
In practical workflows using ODAT-SE, the optimization algorithm is explicitly specified by the variable \texttt{algorithm\_name}, and the algorithm's hyperparameters like the number of Monte Carlo samplings are described using a dictionary representation equivalent to the \texttt{[algorithm]} section of the ODAT-SE input file.
PHYSBO provides a utility function,
\texttt{physbo.search.optimize.odatse.default\_alg\_dict},
which generates a template dictionary with default hyperparameters for the selected algorithm.
For example, a dictionary representing the default settings for the replica exchange Monte Carlo method can be obtained as
\begin{verbatim}
odatse_alg_dict = physbo.search.optimize.odatse.default_alg_dict(
    min_X=min_X, max_X=max_X, algorithm_name="exchange"
)
\end{verbatim}
and users can modify parameters as ones want.
This design enables seamless switching between optimization strategies by modifying only the value of \texttt{algorithm\_name}, while keeping the search space definition unchanged.
Once the paremeter dictionary is prepared, an optimizer is initialized as
\begin{verbatim}
optimizer = physbo.search.optimize.odatse.Optimizer(odatse_alg_dict)
\end{verbatim}
and then passed to a range policy in the same way of the random sampling.

\subsection{Numerical Example}
\begin{figure}[t]
  \centering

  \begin{overpic}[width=0.32\linewidth]{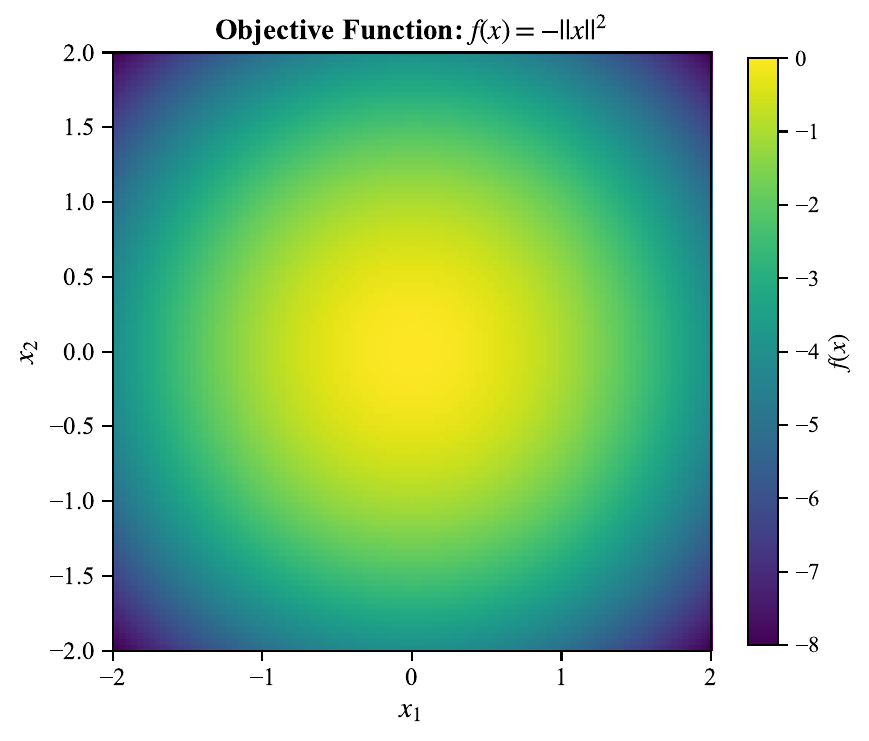}
    \put(3,92){\large\textbf{(a)}}
  \end{overpic}\hfill
  \begin{overpic}[width=0.32\linewidth]{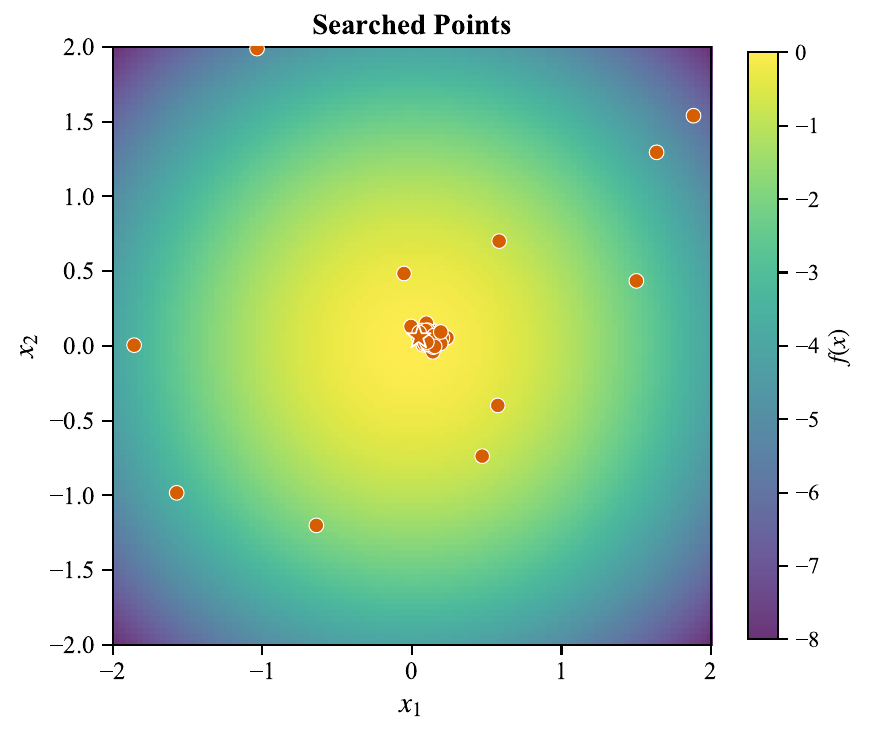}
    \put(3,92){\large\textbf{(b)}}
  \end{overpic}\hfill
  \begin{overpic}[width=0.32\linewidth]{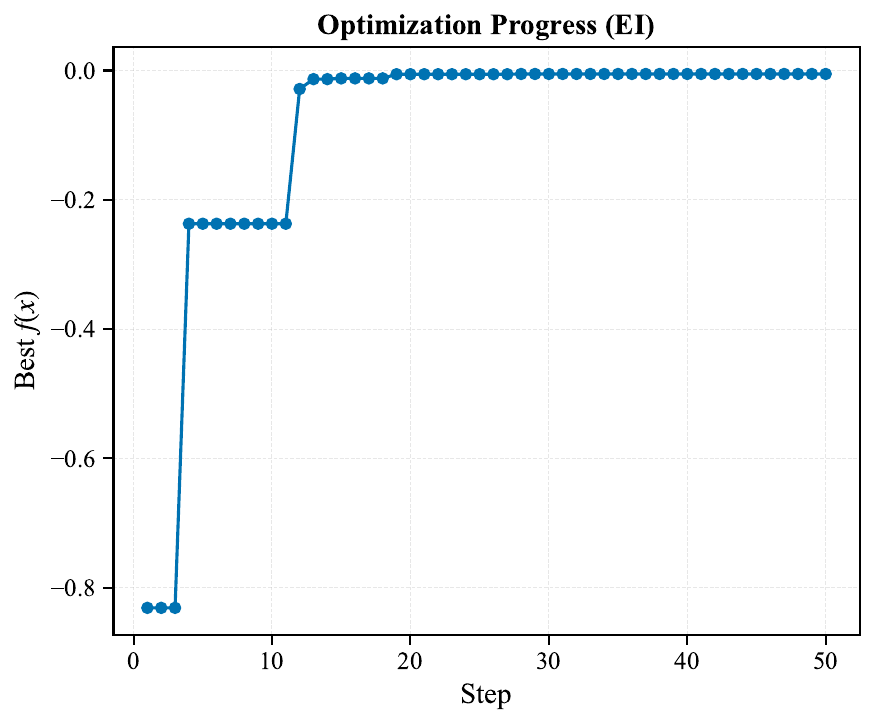}
    \put(3,92){\large\textbf{(c)}}
  \end{overpic}

  \caption{Bayesian optimization in a continuous parameter space using the range
  policy of PHYSBO.
  (a) Objective function landscape
  $f(\boldsymbol{x})=-\lVert\boldsymbol{x}\rVert^2$ over
  $[-2,2]\times[-2,2]$.
  (b) Locations of evaluated points during optimization, showing adaptive
  concentration near the optimum.
  (c) Optimization progress in terms of the best-so-far objective value using
  the expected improvement (EI) acquisition function.}
  \label{fig:range_overview}
\end{figure}

We demonstrate Bayesian optimization in a continuous parameter space using the range policy of PHYSBO, following the implementation shown in the accompanying sample code.
As a simple yet illustrative example, we consider the optimization of a smooth two-dimensional objective function
\begin{equation}
f(\boldsymbol{x}) = -\lVert \boldsymbol{x} \rVert^2,
\qquad
\boldsymbol{x} = (x_1, x_2) \in \mathbb{R}^2,
\end{equation}
which has a unique maximum at $\boldsymbol{x} = (0,0)$ and represents a typical single-basin energy landscape encountered in physics and materials science.
The objective function landscape over the domain $[-2,2]\times[-2,2]$ is shown in Fig.~\ref{fig:range_overview} (a).
The function value decreases smoothly away from the origin, providing a well-behaved test case for continuous Bayesian optimization.

The continuous search space is specified by lower and upper bounds for each parameter, and the range policy is initialized accordingly.
An initial exploration phase is performed using random sampling within the specified parameter ranges.
Subsequently, Bayesian optimization is applied using the expected improvement (EI) acquisition function, which is optimized by random sampling in the continuous parameter space.

Figure~\ref{fig:range_overview} (c) shows the optimization progress in terms of the best objective value found so far.
After an initial rapid improvement during the random sampling phase, the Bayesian optimization stage efficiently refines the search and quickly converges toward the global optimum at $f(\boldsymbol{x}) = 0$.
This behavior illustrates the effectiveness of acquisition-driven search in continuous spaces without discretization.
Figure~\ref{fig:range_overview} (b) visualizes the locations of the evaluated points in the parameter space.
While the initial random samples are broadly distributed, subsequent evaluations are concentrated near the optimum.
This demonstrates how Bayesian optimization adaptively focuses sampling in promising regions of the continuous search space.
Together, these results highlight the ability of PHYSBO to perform efficient Bayesian optimization directly in continuous parameter spaces using the range policy, avoiding the need for explicit discretization and enabling scalable optimization in higher-dimensional settings.

\section{Conclusion}
In this paper, we have presented the major updates introduced in PHYSBO version 3, with a particular emphasis on usability, portability, and practical deployment in real-world physics and materials research. Rather than focusing on the introduction of fundamentally new optimization algorithms, PHYSBO version 3 represents a systematic refinement of the library to better support research workflows.

The updates described in this work address several long-standing challenges encountered in practical Bayesian optimization. Improvements in computational performance and scalability enable stable use in large-scale optimization problems. Enhanced support for multi-objective optimization allows users to handle realistic research scenarios involving competing objectives. The introduction of range-based policies extends PHYSBO to continuous-variable optimization, improving flexibility. In addition, the removal of environment-dependent components, compatibility with modern numerical libraries such as NumPy 2, and the transition to a more permissive software license collectively improve portability, sustainability, and ease of integration.

Taken together, these changes reposition PHYSBO from a specialized optimization tool to a more general research infrastructure for Bayesian optimization. By lowering technical and organizational barriers, PHYSBO version 3 enables researchers to more easily incorporate Bayesian optimization into computational and experimental studies, facilitating rapid iteration, reproducibility, and collaboration. Notably, the removal of the Cython dependency significantly enhances accessibility for Windows users. Given that many experimental instruments operate on Windows environments, this update facilitates the transition of Bayesian optimization from purely cyber-space simulations to physical-world experimental research. These improvements ensure that PHYSBO is more easily integrated into modern AI-robotic research, positioning it to support future developments in automated and data-driven scientific discovery.

\section{Acknowledgement}
We wish to thank Naoki Kawashima and Tsuyoshi Ueno for their highly useful discussions. PHYSBO was developed under the support of the “Project for advancement of software usability in materials science”~\cite{Yoshimi31122025} in fiscal years 2020 and 2025 by the Institute for Solid State Physics, University of Tokyo. This work is partially supported by JSPS KAKENHI Grant Nos 25H01403 and 25K22013. Additional support was provided by the Ministry of Education, Culture, Sports, Science, and Technology (MEXT) through the Data Creation and Utilization Type Material Research and Development Project (JPJ010337, Grant Number: JPMXP1122683430 and JPMXP1121467561).






\bibliographystyle{elsarticle-num}

\end{document}